\documentclass{article}
\usepackage{spconf,amsmath,graphicx}

\title{Deep Factorization for Speech Signal}
\name{Lantian Li, Dong Wang, Yixiang Chen, Ying Shi, Zhiyuan Tang, Thomas Fang Zheng\thanks{
This work was supported by the National Natural Science
Foundation of China under Grant No.61633013 / 61371136 and
the National Basic Research Program (973 Program) of China under
Grant No.2013CB329302. A pre-print version was published on arXiv:1706.01777.
Dong Wang is the corresponding author (wangdong99@mails.tsinghua.edu.cn).}}
\address{Center for Speech and Language Technologies, Research Institute of Information Technology \\
Department of Computer Science and Technology, Tsinghua University, Beijing, 100084, China}%
\begin{document}
%
\maketitle
\begin{abstract}
  Various informative factors mixed in speech signals, leading to great difficulty when decoding
  any of the factors.
  An intuitive idea is to factorize each speech frame into individual informative factors, though it turns out to be
  highly difficult. Recently, we found that speaker traits,
  which were assumed to be long-term distributional properties, are actually short-time patterns, and
  can be learned by a carefully designed deep neural network (DNN).
  This discovery motivated a \emph{cascade deep factorization} (CDF) framework that will be presented in this paper.
  The proposed framework infers speech factors in a sequential
  way, where factors previously inferred are used as conditional variables when inferring other factors.
  We will show that this approach can effectively factorize speech signals,
  and using these factors, the original speech spectrum can be recovered with a high accuracy. This factorization and
  reconstruction approach provides potential values for many speech processing tasks, e.g., speaker recognition and
  emotion recognition, as will be demonstrated in the paper.

\end{abstract}
\section{Introduction}

Speech signals involve rich information,
including linguistic content, speaker trait, emotion, channel and background noise, etc.
Researchers have worked for several decades to
decode these information, leading to
a multitude of speech information processing tasks, including automatic speech recognition (ASR)
and speaker recognition (SRE)~\cite{benesty2007springer}. After a long-term research, some tasks have been addressed pretty well, at least when
a large amount of data is available, e.g., ASR and SRE; while others remain difficult, e.g., automatic emotion recognition (AER)~\cite{el2011survey}.

A major difficulty of speech processing resides in the fact that multiple informative factors are intermingled together,
and so whenever we decode for a particular factor, all other factors contribute as uncertainties. An intuitive idea
to deal with the information blending is to factorize the speech signal into individual informative factors at the frame level.
However, it turns out to be highly difficult, due to at least two reasons: Firstly, the way that these factors are mixed
is unclear and seems highly complex; Secondly, and perhaps more fundamentally, some major factors, particularly the speaker trait,
behaves as long-term distributional properties rather than short-time patterns. It has been partly demonstrated by
the fact that most of the successful speaker recognition approaches (e.g., JFA~\cite{Kenny07}, i-vector~\cite{dehak2011front})
rely on statistical models that retrieve speaker vectors based multiple frames (segments). Therefore, there is a wide suspicion that
speech signals are short-time factorizable.

Fortunately, our recent study showed that speaker traits are largely short-time spectral
patterns, and a carefully designed deep neural network can learn to extract these patterns
at the frame level~\cite{li2017deep}.
The following studies demonstrated that the frame-level deep speaker features are highly generalizable: they
work well with voices of trivial events, such as laugh and cough that are as short as 0.3 seconds~\cite{zhang2017speaker}; and they
are robust against language mismatch~\cite{li2017cross}. The short-time property of speaker traits suggests
that speech signals are possibly short-time factorizable, as it has been known that another major speech factor,
the linguistic content, is also short-time identifiable~\cite{hinton2012deep}.

In this paper, we present a \emph{cascaded deep factorization}
(CDF) approach to obtain such factorization. By this approach,
the most significant factors are inferred firstly, and other less significant factors are inferred
subsequently on the condition of the factors that have already been inferred.
Our experiments on a speaker recognition task and an emotion recognition task demonstrated that the CDF-based factorization
is highly effective. Furthermore, we show that the original speech signal can be reconstructed from
the CDF-derived factors pretty well.

\section{Speaker feature learning}
\label{sec:speaker}

In the previous study~\cite{li2017deep}, we presented a CT-DNN structure that can learn speaker features
at the frame level. The network consists of a convolutional (CN)
component and a time-delay (TD) component.
The CN component comprises two CN layers, each followed by a max-pooling layer.
The TD component comprises two TD layers, each followed by a P-norm layer.
The output of the second P-norm layer is projected into a feature layer. The
activations of the units of this layer, after length normalization, form the speaker feature of the input speech frame.
During model training, the feature layer is fully connected to an output layer whose units correspond to
the speaker identities in the training data. The training is performed to optimize the cross-entropy objective
that aims to discriminate the training speakers based on the input frames.
We demonstrated that the speaker feature inferred by the CT-DNN structure is
highly speaker-discriminative~\cite{li2017deep}, and speaker traits are largely short-time
spectral patterns and can be identified at the frame level.


\section{Cascaded deep factorization}

The ASR research has demonstrated that the linguistic content can be individually
inferred by a DNN at the frame level~\cite{yu2014automatic}, and the deep speaker feature
learning method described in the previous section demonstrated that speaker traits can be
also identified by a very short segment.
We denote this single factor inference method based on deep neural models by \emph{individual deep factorization} (IDF).
The rationality of the IDF method is two-fold: Firstly, the target factor (linguistic content or speaker trait) is
sufficiently significant in speech signals; Secondly, a large amount of training data is available.
It is the large-scale supervised learning that picks up the most task-relevant factors by leveraging the
power of DNNs in feature learning. For factors that are less significant and/or without much
training data, however, IDF is not applicable. Fortunately, the successful inference of the linguistic
and the speaker factors may significantly simplify the inference of those `not so prominent' factors.
This motivated a cascaded deep factorization (CDF) approach: firstly we infer a particular factor
by IDF, and then use this factor as a conditional variable to infer the second factor, and so on.
Finally, the speech signal will be factorized into a set of individual factors, each corresponding to a
particular task.

\begin{figure}[htb]
    \centering
    \includegraphics[width=0.99\linewidth]{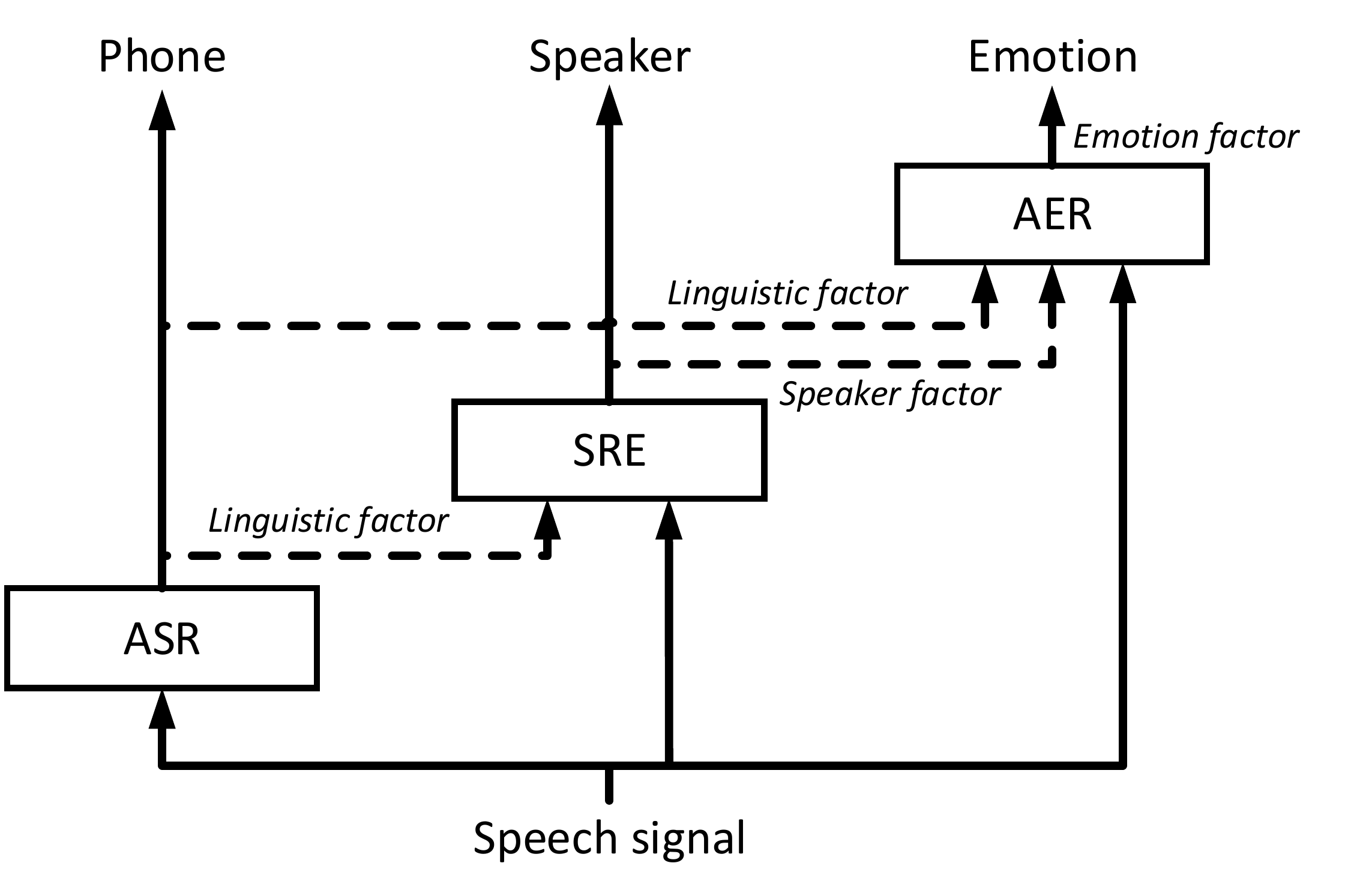}
    \caption{The cascaded deep factorization approach applied to factorize emotional speech into
    three factors: linguistic, speaker and emotion. }
    \label{fig:cascade}
\end{figure}

To demonstrate this concept, we apply the CDF approach to factorize emotional speech signals into three factors:
linguistic, speaker and emotion. Fig.~\ref{fig:cascade} illustrates the architecture. Firstly
an ASR system is trained using word-labelled speech data.
The frame-level linguistic factor, which is in the form of phone posteriors in our study, is produced
from the ASR system, and is concatenated with the raw feature (Fbanks) to train an SRE system.
This SRE system is used to produce the frame-level speaker factor, as discussed in the previous section.
The linguistic factor and the speaker factor are finally concatenated with the raw
feature to train an AER system, by which the emotion factor is produced from the last hidden layer.

The CDF approach is fundamentally different from the conventional factorization approach,
e.g., JFA~\cite{Kenny07}: (1) CDF is frame-level while conventional methods are segment-level; (2)
CDF relies on discriminative training while conventional factorization methods
rely on maximum likelihood estimation; (3) CDF infers factors sequentially and so can use data with
partial labels (e.g., only speaker labels), while conventional approaches infer factors jointly
so can only use full-labelled data; (4) CDF are based on DNNs that are deep, non-linear and non-Gaussian,
while most conventional approaches are based on models that are shallow, linear and Gaussian.

We highlight that more complex model structures are possible to conduct the factorization, e.g.,
with the collaborative learning architecture~\cite{Tang2017Collaborative}. However, the CDF framework is
consistent with a cascaded convolution view for speech signals, i.e., speech signals
are produced by convolving informative factors sequentially, from linguistic parts to non-linguistic
parts, as mentioned in~\cite{fujisaki1998communication, sagisaka2012computing}.

\section{Spectrum reconstruction}
\label{sec:recovery}

\begin{figure}[htb]
    \centering
    \includegraphics[width=0.99\linewidth]{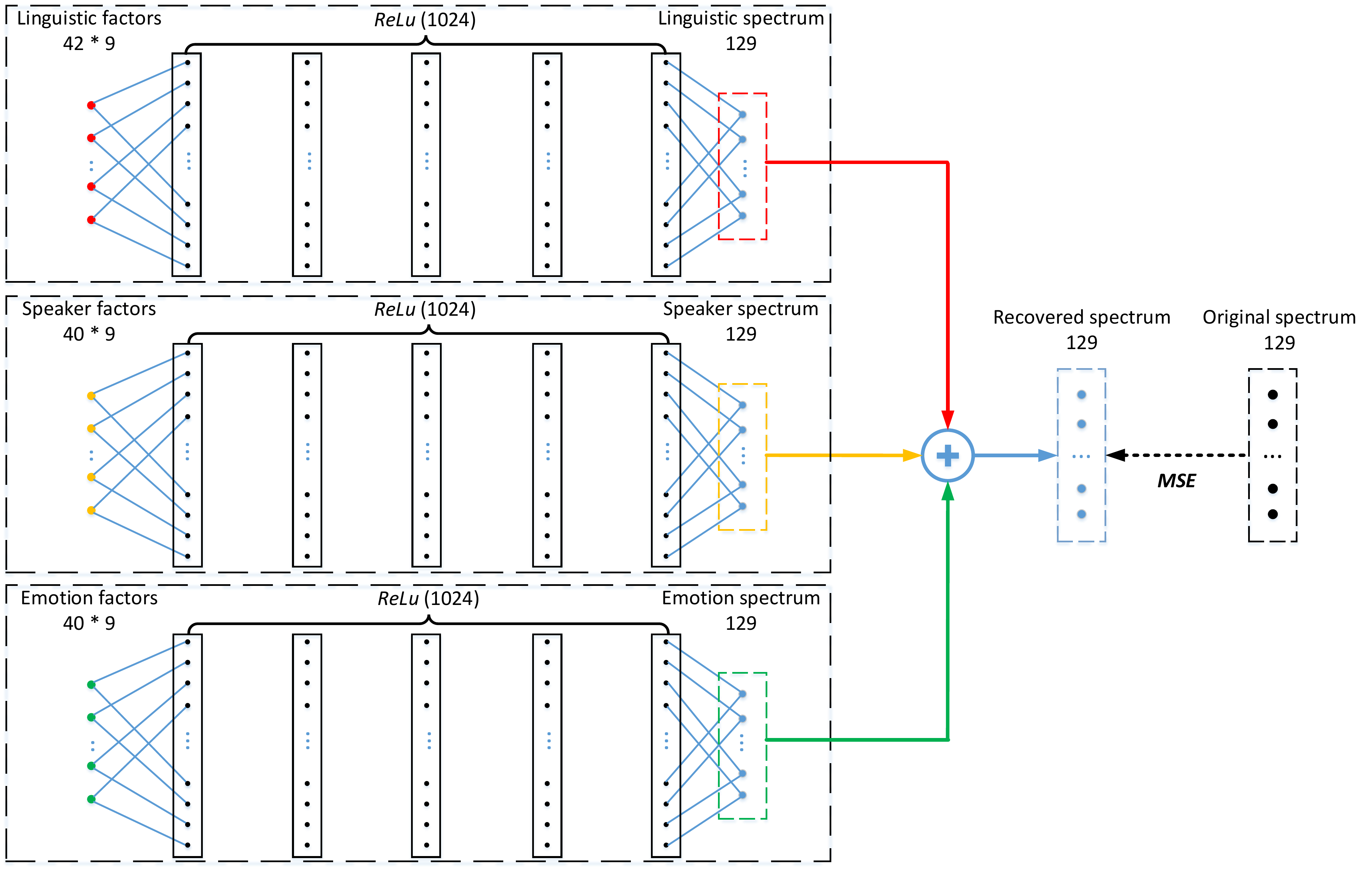}
    \caption{The architecture for cascaded convolutional spectrum reconstruction.}
    \label{fig:recovery}
\end{figure}

An interesting property of the CDF-inferred factors is that they can be used to recover the original
speech. Define
the linguistic factor $q$, the speaker factor $s$, and the emotion factor $e$.
For each speech frame, we try to use these three factors to
recover the spectrum $x$. According to the cascaded convolution view, the reconstruction is written in the form:

\begin{equation}
\label{eq:rec}
   ln (x) = ln \{f(q)\} + ln \{g(s)\} + ln \{h(e)\} + \epsilon
\end{equation}

\noindent where $f$, $g$, $h$ are the non-linear recovery function for $q$, $s$ and $e$ respectively,
each implemented as a DNN. $\epsilon$ represents the residual which is assumed to be Gaussian.
This reconstruction is illustrated in Fig.~\ref{fig:recovery}, where all the spectra are in the log domain.
Note that $q$, $s$, $e$ are all inferred from Fbanks rather than the original spectra, but they can still recover
the original signal pretty well, as will be seen in Section~\ref{sec:exp}.

\section{Related work}


The CDF approach shared a similar motivation as the phonetic DNN i-vector
approach~\cite{lei2014novel,Kenny14}: both utilize the phonetic factor to support inference
of other factors. The difference is that CDF is a neural model and retrieves frame-level features,
while phonetic DNN i-vector is a probabilistic model and retrieves utterance-level representations.
CDF is also related to multi-task learning~\cite{caruana1997multitask} and transfer
learning~\cite{pan2010survey,wang2015transfer}, where multiple tasks are used to regularize the
training~\cite{Tang2017Collaborative,qian2016neural,li2015modeling}. Compared to these
methods, the CDF approach focuses on explaining variabilities of speech signals, rather than
learning models. Finally, the CDF approach is also related to auto-encoder (AE) that can be
regarded as an unsupervised factorization. Compared to AE, CDF is a supervised learning, and
the learned factors are inherently task-oriented.

\section{Experiment}
\label{sec:exp}

\subsection{Database}

Three databases are used in our experiment, as presented below.
All the speech signals are down-sampled to 8k Hz to ensure data consistency.

\textbf{ASR database}: The \emph{WSJ} database was used to train the ASR system. The training data
is the official \emph{train\_si284} dataset, composed of $282$ speakers and $37,318$ utterances.
The test set contains three datasets (\emph{devl92, eval92 and eval93}),
including $27$ speakers and $1,049$ utterances in total.

\textbf{SRE database}: The \emph{Fisher} database was used to train the SRE systems. The training set
consists of $2,500$ male and $2,500$ female speakers, with $95,167$ utterances in total.  Each
speaker has about $120$ seconds of speech signals. It was used for training the UBM, T-matrix and
PLDA models of an i-vector baseline system, and the DNN model described in Section~\ref{sec:speaker}.
The evaluation set consists of $500$ male and $500$ female speakers, $82,990$ utterances in total. The speakers
of the training set and the evaluation set are not overlapped. For each speaker, $10$ utterances (about $30$ seconds in total)
are used for enrollment and the rest for test.

\textbf{AER database}: The \emph{CHEAVD} database~\cite{bao2014building} was used to train the AER systems.
This database was selected from Chinese movies and TV programs and was used as the standard database for the
multimodal emotion recognition challenge (MEC 2016)~\cite{li2016mec}.
There are $8$ emotions in total: Happy, Angry, Surprise, Disgust, Neutral, Worried, Anxious and Sad.
The training set contains $2,224$ utterances and the evaluation set contains $628$ utterances.

\subsection{ASR baseline}

We first build a DNN-based ASR system using the WSJ database. This system will be used to produce the linguistic factor
in the following CDF experiments. The Kaldi toolkit~\cite{povey2011kaldi} is used to train the DNN model,
following the Kaldi WSJ s5 nnet recipe. The DNN structure consists of $4$ hidden layers, each containing $1,024$ units.
The input feature is Fbanks, and the output layer discriminates $3,383$ GMM pdfs. With the official 3-gram language model,
the word error rate (WER) of this system is $9.16$\%. The linguistic factor is represented by
the $42$-dimensional phone posteriors, derived from the output of the ASR DNN.

\subsection{SRE by IDF and CDF}

We build three SRE systems: an i-vector/PLDA system~\cite{dehak2011front,Ioffe06} to represent the conventional statistical  model approach;
an IDF d-vector system that follows the CT-DNN architecture,
where only the raw features (Fbanks) comprise the input; and a CDF d-vector system that
follows the CDF spirit and the linguistic factors produced by the ASR system are used as
additional input to the CT-DNN architecture.

For the i-vector system, the UBM is composed of $2,048$
Gaussian components, and the i-vector dimension is set to $400$. The system is trained
following the Kaldi SRE08 recipe. For the d-vector systems, the frame-level speaker
features are of $40$ dimensions, and the utterance-level d-vector is derived as
an average of the frame-level features within the utterance.
More details of the d-vector systems can be found in~\cite{li2017deep}.

We report the results on the identification task, though similar observations were obtained
on the verification task.
In the identification task, one matched speaker (\emph{Top-1}) is identified given a
test utterance. With the i-vector (d-vector) system, each enrolled speaker is represented by
the i-vector (d-vector) of their enrolled speech, and the i-vector (d-vector) of the test speech
is compared with the i-vectors (d-vectors) of the enrolled speakers, finding the speaker whose
enrollment i-vector (d-vector) is nearest to that of the test speech.
For the i-vector system, the popular PLDA model~\cite{Ioffe06} is used to measure the similarity between
i-vectors; for the d-vector system, the simple cosine distance is used.

The results in terms of the Top-1 identification rate (IDR) are shown in Table~\ref{tab:id-short}.
In this table, `C(30-20f)' means the test condition where the duration of the enrollment speech is $30$ seconds, while the test
speech is $20$ frames. Note that $20$ frames is just the length of the effective context window of the
speaker CT-DNN, so only one single frame of speaker feature is used in this condition. From these results,
it can be observed that the d-vector system performs much better than the i-vector baseline, particularly with
very short speech segments. Comparing the IDF and CDF results, it can be seen that the CDF approach that
involves phonetic knowledge as the conditional variable greatly improves the d-vector system in the short speech
segment condition.

    \begin{table}[htb!]
    \begin{center}
      \caption{The \textbf{\emph{Top-1}} IDR(\%) results on the short-time speaker identification with the i-vector and two d-vector systems.}
      \label{tab:id-short}
        \scalebox{0.9}{
          \begin{tabular}{|c|l|c|c|c|c|}
            \hline
            \multicolumn{2}{|c|}{}                 &\multicolumn{3}{c|}{IDR\%}\\
            \hline
               Systems              &  Metric    &   C(30-20f) &    C(30-50f)   &   C(30-100f) \\
           \hline
              i-vector               &    PLDA     &   5.72    &    27.77      &    55.06      \\
               d-vector (IDF)             &    Cosine   &   37.18   &    51.24 &    \textbf{65.31}      \\
               d-vector (CDF)        &    Cosine   & \textbf{47.63}   & \textbf{57.72}      &    64.45      \\
           \hline
          \end{tabular}
          }
      \end{center}
   \end{table}

\subsection{AER by CDF}

This section applies the CDF approach to an emotion recognition task.
For that purpose, we first build a DNN-based AER baseline. The DNN model consists of $6$
hidden layers, each containing $200$ units. After each layer, a P-norm layer reduces the dimensionality
from $200$ to $40$. The output layer comprises $8$ units, corresponding to the number of emotion classes
in the CHEAVD database. This DNN model produces frame-level emotion posteriors. The utterance-level posteriors
are obtained by averaging the frame-level posteriors, by which the utterance-level emotion decision is achieved.

Three CDF configurations are investigated, according to which factor is used as the conditional:
the linguistic factor (+ ling.), the speaker factor (+ spk.) and both (+ ling. \& spk.).
The results are evaluated in two metrics:
the identification accuracy (ACC) that is the ratio of the correct identification on all emotion categories;
the macro average precision (MAP) that is the average of the ACC on each of the emotion category.

\begin{table}[thb!]
  \caption{\label{tab:cdf}{Accuracy (ACC) and macro average precision (MAP) of the AER systems.}}
  \centerline{
  \scalebox{0.8}{
    \begin{tabular}{|l|c|c|c|c|c|}
      \hline
                       & \multicolumn{4}{|c|}{Training set} \\
      \hline
                       & ACC\% (fr.) & MAP\% (fr.) & ACC\% (utt.) & MAP\% (utt.) \\
      \hline
      Baseline         & 74.19 & 61.67 & 92.27  & 83.08               \\
      +ling.           & 86.34 & 81.47 & 96.94  & 96.63               \\
      +spk.            & 92.56 & 90.55 & 97.75  & 97.16               \\
      +ling. \& spk.      & \textbf{94.59} & \textbf{92.98} & \textbf{98.02}  & \textbf{97.34}   \\
      \hline
      \hline
                   & \multicolumn{4}{|c|}{Evaluation set}\\
      \hline
                   & ACC\% (fr.) & MAP\% (fr.) & ACC\% (utt.) & MAP\% (utt.)  \\
      \hline
      Baseline     & 23.39 & 21.08 & 28.98  & 24.95  \\
      +ling.       & 27.25 & 27.68 & \textbf{33.12}  & \textbf{33.28}  \\
      +spk.        & 27.18 & 28.99 & 32.01  & 32.62  \\
      +ling. \& spk.  & \textbf{27.32} & \textbf{29.42} & 32.17  & 32.29  \\
      \hline
   \end{tabular}
  }
  }
\end{table}

The results on the training set are shown in Table~\ref{tab:cdf}, where the ACC and MAP values
on both the frame-level (fr.) and the utterance-level (utt.) are reported.
It can be seen that with the conditional factors involved, the ACC and MAP values on
the training set are significantly improved, and the speaker factor seems to provide more contribution.
This improvement on training accuracy demonstrates that with the conditional factors considered,
the speech signal can be \emph{explained} much better.

The results on the evaluation set are also shown in Table~\ref{tab:cdf}. Again, we observe a clear advantage
with the CDF training. Note that involving the two factors does not improve the
utterance-level results. This should be attributed to the fact that the DNN models are trained
using frame-level data, so may be not fully consistent with the metric of the utterance-level test.
Nevertheless, the superiority of the multiple conditional factors can be seen clearly from the frame-level
metrics. We note that the discrepancy between the results on the training set and the evaluation
set is not due to over-fitting caused by the CDF approach; it simply reflects the mismatch between
the training set and evaluation set, hence the difficulty of the task itself.
Actually, the results shown here are highly competitive: it beats the MEC 2016 baseline~\cite{li2016mec}
by a large margin, even though we used the 8k Hz data rather than the original 16k Hz data.

\subsection{Spectrum reconstruction}

\begin{figure}[htb]
    \centering
    \includegraphics[width=1\linewidth]{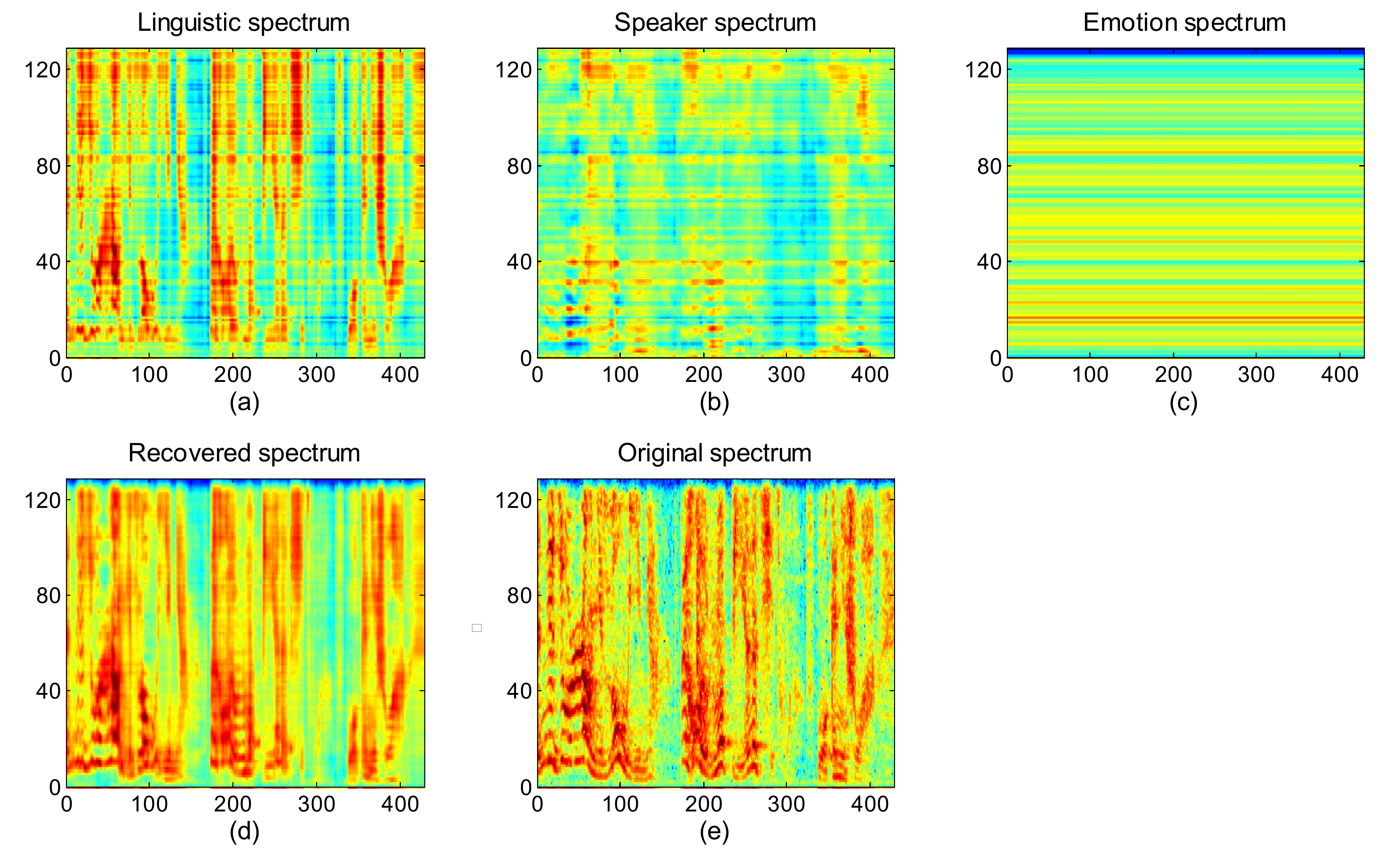}
    \caption{An example of spectrum reconstruction from the linguistic, speaker and emotion factors.}
    \label{fig:demo}
\end{figure}

In the last experiment, we use the linguistic factor, speaker factor and emotion factor to reconstruct
the original speech signal, using the convolutional model shown in Fig.~\ref{fig:recovery}.
The model is trained using the CHEAVD database. During the training processing, the averaged frame-level
reconstruction loss (square error)
on the validation set is reduced from $15285.70$ to $192.50$, and the loss on the evaluation set with the trained
model is $196.56$.
Fig.~\ref{fig:demo} shows a reconstruction example, where the utterance is selected from the test set of
the CHEAVD database. It can be seen
that these three factors (linguistic, speaker, emotion) can reconstruct the original spectrum pretty well.
Listening tests show that the reconstruction quality is rather good and it is hard for human listeners
to tell the difference between the reconstructed speech and the original speech.
More examples can be found in the project web site \emph{http://project.cslt.org}.
The success of this `deep reconstruction' indicates that the CDF approach not only factorizes speech signals into
task-oriented informative factors, but also preserves most of the information of the speech
during the factorization. Moreover, it demonstrates that the cascaded convolution view (Eq.~\ref{eq:rec})
is largely correct. This essentially provides a new vocoder that decomposes speech signals into a sequential convolution
of task-oriented factors, which is fundamentally different from the classical source-filter model.

\section{Conclusions}

This paper presented a cascaded deep factorization (CDF)
approach to factorize speech signals into individual task-oriented informative factors.
Our experiments demonstrated that speech signals can be well factorized at the frame level by the CDF approach,
and speech signals can be largely reconstructed using deep neural models
from the CDF-derived factors. Moreover, the results on the emotion recognition task demonstrated
that the CDF approach is particularly valuable for learning and inferring less significant factors
of speech signals.



\bibliographystyle{IEEEbib}
\bibliography{refs}

\end{document}